\documentclass[11pt]{article}

\usepackage{amsfonts}
\usepackage{amsmath}
\usepackage{amssymb}
\usepackage{amsthm}

\usepackage{graphicx}

\theoremstyle{plain}

\newtheorem{theorem}{Theorem}

\newtheorem{corollary}{Corollary}
\newtheorem{proposition}{Proposition}

\theoremstyle{definition}

\usepackage[numbers,sort]{natbib}

\begin{document}
\title{On the spread of epidemics in a closed heterogeneous population}

\author{Artem S Novozhilov\footnote{Tel.:+1 (301) 402-4146;
              email: novozhil@ncbi.nlm.nih.gov}\\\textit{\normalsize National Institutes of Health, 8600 Rockville Pike, Bethesda, MD 20894, USA}}

\date{}

\maketitle

\begin{abstract}
Heterogeneity is an important property of any population
experiencing a disease. Here we apply general methods of the
theory of heterogeneous populations to the simplest mathematical
models in epidemiology. In particular, an SIR
(susceptible-infective-removed) model is formulated and analyzed
for different sources of heterogeneity. It is shown that a
heterogeneous model can be reduced to a homogeneous model with a
nonlinear transmission function, which is given in explicit form.
The widely used power transmission function is deduced from a
heterogeneous model with the initial gamma-distribution of the
disease parameters. Therefore, a mechanistic derivation of the phenomenological model, which mimics reality very well, is provided. The equation for the final size of an epidemic
for an arbitrary initial distribution is found. The implications
of population heterogeneity are discussed, in particular, it is pointed out that usual moment-closure methods can lead to erroneous conclusions if applied for the study of the long-term behavior of the model.
\paragraph{Keywords:} SIR model, heterogeneous populations, transmission function
\end{abstract}

\section{Introduction}

Mathematical modeling of epidemics arguably began with the pioneer
work of Ross \cite{Ross1910} and developed considerably ever since
(e.g., \cite{Anderson1991,Diekmann2000}). Especially significant
contribution was made by Kermack and McKendrick
\cite{Kermack1927}, two students of Ross, who considered the
situation of microparasite infection, where contacts between
individuals are made according to the law of mass-action, all
individuals are identical, the population is closed, and the
population size is large enough to apply a deterministic
description (for a brief review see \cite{Diekmann1993}).
Additionally, if it is assumed that the individuals are infected
for an exponentially distributed period of time, then a usual SIR
model in the form of ordinary differential equations (ODEs) can be
written down for the sizes of susceptible, infective and removed
classes.

Since the work of Kermack and McKendrick a great deal of
mathematical models were suggested that relax one or more
assumptions that led to the Kermack-McKendrick model. A
substantial part of this work was devoted to incorporate
heterogeneity into the mathematical models, which is also the main
subject of the present text. In what follows we retain the
assumptions of random mixing, no inflow of susceptible or infected
hosts, exponentially distributed infectious period, and validity
of deterministic description. Specifically, we will look into
heterogeneity in disease parameters (such as susceptibility to a
disease); disease parameters are considered as an inherent and
invariant property of individuals, whereas the parameter values
can vary between individuals. We analyze heterogeneity that was
termed ``parametric'' by Dushoff \cite{Dushoff1999} not addressing
important topics of heterogeneity mediated by a structured
variable, such as explicit space or age structure.

The most common way to take into account parametric heterogeneity
is to divide population into groups \cite{Dushoff1995,Hsu2002,
Hyman2005,Jacquez1995}. An important disadvantage of the subgroup
approach is that heterogeneity within a group cannot be
incorporated. Another approach, which we also pursue, is to
consider the population as having a continuous distribution (see,
e.g.,
\cite{Diekmann1990,Diekmann1993,Diekmann2000,Dwyer1997,Dwyer2000,Dwyer2002,Nikolaou2006,Veliov2005})
or a very large number of subgroups as it was done by May et al.
\cite{May1988} (eventually they used  a continuous
gamma-distribution).

Our approach to formulate mathematical models is close to that
applied in, e.g., \cite{Dwyer1997}, where known experimental data
forced the authors to take into account heterogeneity among hosts
in their susceptibility to the virus among other key details, and
a simple SIR model was adjusted to account for new information.
The major novelty of the present text is to introduce the well
developed theory of heterogeneous populations into the
epidemiological modeling. Using simple models we are able to
obtain known results with less effort, and, more importantly,
produce new analytical results. In particular, we show that any
heterogeneous SIR model can be reduced to a homogeneous one with a
nonlinear transmission function, and present the exact form of
this function. It turns out that widely applied nonlinear transmission function in the form of power relationship, $\beta S^pI^q$, can be a consequence of the intrinsic heterogeneity in susceptibility and infectivity parameters. For a heterogeneous SIR model the equation of the final epidemic size is found. The explicit form of the final epidemic size emphasizes and illustrates the fact that the goal to model the evolution of a heterogeneous population for a long time can be accomplished only if the exact initial distribution is available. Any moment-closure methods may lead to erroneous estimates.

Our paper is organized as follows. In Section 2 we formulate the
basic models and discuss various assumptions that might lead to
them. In Section 3 we review the necessary analytical tools from
the theory of heterogeneous populations. In Section 4 a
homogeneous model that is equivalent to the heterogeneous one is
explicitly constructed, and it is shown that the former has a
nonlinear transmission function; moreover, the well known power
transmission function is shown to be a consequence of the initial
gamma-distribution. In Section 5 the influence of heterogeneity on
the disease course is studied for an SIR model with distributed
susceptibility, in particular, the final size of an epidemic is
found for an arbitrary initial distribution. Section 6 devoted to
discussion and conclusions. Finally, in Appendix we collect the
definitions of the probability distributions used throughout the
text together with some auxiliary facts from the general theory of
heterogeneous models.

\section{The basic models with population heterogeneity}
\subsection{The model with distributed susceptibility}\label{s1}

Suppose that each individual of a (sub-)population has its own
value of a certain trait (which can be, e.g., susceptibility to a
particular disease, social behavior, infectivity, or a hereditary
attribute) that describes his or her invariant property and has a
marked influence on the disease course; that is, the key parameters
that determine disease evolution depend on the trait values and we
can speak of the trait distribution or the parameter distributions
(in general, we speak of a distribution when no ambiguity is
expected). The trait value remains unchanged for any given
individual during the time period we are interested in, but varies
from one individual to another. Any changes of the mean, variance
and other characteristics of the trait distribution in the
population (or the parameter distributions) are caused only by
variation of population structure. Such situation is obviously
closer to the reality then the usual assumption of the population
uniformity in the SIR-like models described by ODEs.

For the moment we assume that the susceptible subpopulation
is heterogeneous, and denote $s(t,\omega)$ the density of
susceptible individuals at time $t$ having trait value $\omega$
(i.e., the size of subpopulation of susceptible hosts having trait
values in the range from $\omega$ to $\omega+d\omega$ is
approximately equal to $s(t,\omega)d\omega$, and the total size of
the susceptibles $S(t)=\int_{\Omega}s(t,\omega)d\omega$, where
$\Omega$ is the set of trait values). Assuming that the
subpopulation of infectives is homogeneous (under the modeling
situation), the total size of the population is constant, the
disease course keeps within the simplest situation
``susceptible''$\rightarrow$``infective''$\rightarrow$``removed'',
the contact process is described by the so-called mass-action
kinetics (i.e., the contact rate is proportional to the total
population size $N$
\cite{Diekmann2000,Heesterbeek2005,mccallum2001}), and that the
rate of change in the susceptibles is determined by the
transmission parameter, which is a function of trait values, we
can write down the following equation for the change in the
susceptible subpopulation having trait value $\omega$:
\begin{equation}\label{s1:1}
    \frac{\partial}{\partial
    t}s(t,\omega)=-\beta(\omega)s(t,\omega)I(t),
\end{equation}
where $I(t)$ is the size of the subpopulation of infectives, and
$\beta(\omega)$ incorporates information on the contact rate and
the probability of a successful contact. Hereinafter we assume
that the trait that characterizes susceptible individuals is the
susceptibility to the disease, although it can be assumed that
different individuals have different contact rates (in the latter
case it becomes difficult to interpret the equation for the
infectives, see below).

The change in the infective class, if the length of being
infective is distributed exponentially with the mean time
$1/\gamma$, is given by
\begin{equation}\label{s1:2}
    \frac{d}{dt}I(t)=I(t)\int_{\Omega}\beta(\omega)s(t,\omega)\,d\omega-\gamma
    I(t)=\bar{\beta}(t)S(t)I(t)-\gamma I(t),
\end{equation}
where we used the useful notations
$$
\bar{\beta}(t)=\int_{\Omega}\beta(\omega)p_s(t,\omega)\,d\omega,\quad
p_s(t,\omega)=\frac{s(t,\omega)}{S(t)}\,.
$$
Hence, $\bar{\beta}(t)$ is the mean value of the function
$\beta(\omega)$, and $\omega$ has the probability density function
(pdf) $p_s(t,\omega)$ for any time moment $t$. We need to
supplement the model \eqref{s1:1}-\eqref{s1:2} with the initial
conditions
\begin{equation}\label{s1:3}
    s(0,\omega)=s_0(\omega)=S_0p_s(0,\omega),\quad I(0)=I_0,\quad
    R(0)=0,
\end{equation}
and with the third equation for the removed $dR(t)/dt=\gamma
I(t)$. Here $S_0$ and $I_0$ are the initial sizes of the susceptibles and infectives respectively, and $p_s(0,\omega)$ is the given initial distribution.

The model \eqref{s1:1}-\eqref{s1:3} is the basic model we study in
this paper. This model was formulated from the first principles,
as it was done for conceptually similar models in \cite{Dwyer1997}
for the transmission of virus in gypsy moths, in
\cite{Nikolaou2006} for the effect of antimicrobial agents on
microbial populations, in \cite{May1988} for the spread of HIV in
the human population, and in \cite{Veliov2005} for a class of SIS
models (we note, however, that in the last two examples the
frequency-dependent transmission was employed \cite{mccallum2001},
and the heterogeneity of the contact rates was modelled).

The model \eqref{s1:1}-\eqref{s1:3} can be also deduced from the
general epidemic equation (see \cite{Diekmann2000})
\begin{equation}\label{s1:4}
    \frac{\partial}{\partial
    t}s(t,\omega)=s(t,\omega)\int_{\Omega}\int_{0}^{\infty}A(\tau,\omega,\eta)\frac{\partial}{\partial
    t}s(t-\tau,\eta)\,d\tau d\eta,
\end{equation}
where $A(\tau,\omega,\eta)$ is the expected infectivity of an
individual that was infected $\tau$ units of time ago while having
trait value $\eta$ towards a susceptible with trait value
$\omega$. If we assume that
$A(\tau,\omega,\eta)=\beta(\omega)\exp\left\{-\gamma
\tau\right\}$, and set
$$
I(t)=-\int_{\Omega}\int_0^\infty
\exp\{-\gamma\tau\}\frac{\partial}{\partial
t}s(t-\tau,\eta)\,d\tau d\eta,
$$
after some algebra we obtain \eqref{s1:1}-\eqref{s1:3} (see also
\cite{Diekmann1993}). As a side remark we note that letting
$A(\tau,\omega,\eta)=\beta(\omega)\chi(T-\tau)$, where $\chi(t)$ is the
Heaviside function, we obtain the model studied in~\cite{Dwyer1997}.
\subsection{Model with distributed infectivity}\label{s22}

It can be also assumed that the population of infectives is
heterogeneous. Now let $\beta(\omega)$ be the infectivity of an
individual with trait value $\omega$, and $i(t,\omega)$ be the
density of the infective hosts with trait value $\omega$ at time
moment $t$, $I(t)=\int_{\Omega}i(t,\omega)\,d\omega$. For
simplicity we assume that the susceptible hosts are homogeneous.
The change in the infective subpopulation  should incorporate the
law that specifies which trait value is assigned to a newly
infected individual, and can be described by the following
equation:
\begin{equation}\label{s1:5}
    \frac{\partial}{\partial t}i(t,\omega)=S(t)\int_{\Omega}\psi(\omega,\omega')\beta(\omega')i(t,\omega')\,d\omega'-\gamma i(t,\omega),
\end{equation}
where $\psi(\omega,\omega')$ is the probability that a newly
infected individual gets trait value $\omega$ if infected by an
individual with trait value $\omega'$. The change in the
population $S(t)$ is given by
\begin{equation}\label{s1:6}
    \frac{d}{dt}S(t)=-\bar{\beta}(t)S(t)I(t),
\end{equation}
where now
$\bar{\beta}(t)=\int_{\Omega}\beta(\omega)p_i(t,\omega)\,d\omega$,
and $p_i(t,\omega)=i(t,\omega)/I(t)$.

The need to specify function $\psi(\omega,\omega')$ precludes the
interpretation of the function $\beta(\omega)$ in Section \ref{s1}
as the heterogeneity in the contact rates. It was assumed that the
infectives are all identical, and thus the supposition that an
individual that have had the trait value $\omega$ turns into
another identical infective host is not warranted, at least within
the framework of the simple model \eqref{s1:1}-\eqref{s1:3}.

A variety of choices for the function $\psi(\omega,\omega')$ is
possible, but we especially interested in the particular case when
a newly infected individual gets the same trait value that was
possessed by the individual who passed the infection, namely
$$
\psi(\omega,\omega')=\delta(\omega'-\omega),
$$
where $\delta(\omega)$ is the Dirac delta function (this is
similar to the assumptions made in \cite{Veliov2005}). In this
case the equation \eqref{s1:5} simplifies to the equation, which
is very similar to \eqref{s1:1}:
\begin{equation}\label{s1:7}
    \frac{\partial}{\partial t}i(t,\omega)=\beta(\omega)S(t)i(t,\omega)-\gamma
    i(t,\omega).
\end{equation}

Model \eqref{s1:6}-\eqref{s1:7} is another example of a simple
mathematical model for the spread of an infectious disease in a
closed population with heterogeneities. The list of possible
models can be easily extended. For instance, it is straightforward
to assume that the parameter $\gamma$ is not constant for the
infected individuals, but rather is distributed with a known
initial distribution. In this case the equation for the infectives
takes the form
$$
\frac{\partial}{\partial t}i(t,\omega)=\beta
S(t)i(t,\omega)-\gamma(\omega)
    i(t,\omega),
$$
where $\beta$ is now constant. Another obvious generalization is
to assume that several model parameters are distributed.

Models \eqref{s1:1}-\eqref{s1:3} and \eqref{s1:6}-\eqref{s1:7} are
infinite-dimensional dynamical systems where the evolutionary
operator specifies complex transformations of the initial
distributions. Such models are less amenable to qualitative,
quantitative, or numerical analysis than their finite-dimensional
analogs formulated in terms of ODEs. A usual practice is to
formulate an infinite-dimensional system of ODEs for which some
approximations methods (e.g., assuming that the initial
distribution is close to the normal distribution
\cite{Nikolaou2006}), moment-closure (\cite{Dwyer2002,Dwyer2000}
and \cite{Dushoff1999}), or numerical methods (\cite{Veliov2005})
can be applied. We show below that in some particular cases, when
the analytical form of the heterogeneous models meets certain
requirements, the initial model can be reduced to a
low-dimensional ODE model, which, in turn, can be effectively
analyzed.

\section{The necessary facts from the theory of heterogeneous
populations}\label{s2} To keep the exposition self-contained and for the sake of convenience of references we briefly survey the necessary results
from \cite{Karev2000a,Karev2005a}. We present the results in the
form suitable for our goal noting that more general cases can be
analyzed \cite{Karev2005a}. For the proofs we refer to
\cite{karev2006}, where similar models are considered. Some
additional facts are given in Appendix.

Let us assume that there are two interacting populations whose
dynamics depend on trait values $\omega_1$ and $\omega_2$
respectively. The densities are given by $n_1(t,\omega_1)$ and
$n_2(t,\omega_2)$, and the total population sizes
$N_1(t)=\int_{\Omega_1}n_1(t,\omega)\,d\omega_1$ and
$N_2(t)=\int_{\Omega_2}n_2(t,\omega)\,d\omega_2$. Obviously, more than two populations can be considered, or some populations may be supposed to be homogeneous; we choose two not to be drowned in notations. Assume next that the net reproduction rates of the
populations have the specific form which is presented below:
\begin{equation}\label{s2:1}
    \begin{split}
\frac{\partial}{\partial
    t}n_1(t,\omega_1)&=n_1(t,\omega_1)[f_1(N_1,N_2,\bar{\varphi}_2(t))+\varphi_1(\omega_1)g_1(N_1,N_2,\bar{\varphi}_2(t))],\\
    \frac{\partial}{\partial
    t}n_2(t,\omega_1)&=n_2(t,\omega_2)[f_2(N_1,N_2,\bar{\varphi}_1(t))+\varphi_2(\omega_2)g_2(N_1,N_2,\bar{\varphi}_1(t))],
\end{split}
\end{equation}
where $\varphi_i(\omega_i)$ are given functions,
$\bar{\varphi}_i(t)=\int_{\Omega_i}\varphi_i(\omega_i)p_i(t,\omega_i)\,d\omega_i$
are the mean values of $\varphi_i(\omega_i)$, and
$p_i(t,\omega_i)=n_i(t,\omega_i)/N_i(t)$ are the corresponding
pdfs, $i=1,2$. We also assume that $\varphi_i(\omega_i)$,
considered as random variables, are independent. The system
\eqref{s2:1} plus the initial conditions
\begin{equation}\label{s2:1a}
n_i(0,\omega_i)=N_i(0)p_i(0,\omega_i),\,i=1,2,
\end{equation}
defines, in general, a complex transformation of densities
$n_i(t,\omega_i)$. For the approach to study such systems based on the analysis of abstract differential equations in Banach spaces we refer to
\cite{Ackleh1998,Ackleh1999,Ackleh2000}. Another approach to
analyze models in the form \eqref{s2:1} was suggested in
\cite{Karev2000a}. The latter is more attractive because
eventually one has to deal with systems of ODEs of low dimensions (examples of model analysis are given in \cite{Karev2005,Karev2003,karev2006,Novozhilov2004}).

Let us denote
$$
M_i(t,\lambda)=\int_{\Omega_i}e^{\lambda\varphi_i(\omega_i)}p_i(t,\omega_i)\,d\omega_i,\quad
i=1,2,
$$
the moment generating functions (mgfs) of the functions
$\varphi_i(\omega_i)$, $M_i(0,\lambda)$ are the mgfs of the
initial distributions, $i=1,2$, which are given.

Let us introduce auxiliary variables $q_i(t)$ as the solutions of
the differential equations
\begin{equation}\label{s2:2}
    dq_i(t)/dt=g_i(N_1,N_2,\bar{\varphi}_{i+1}(t)),\quad
    q_i(0)=0,\quad
    i=1,2,
\end{equation}
where indexes that exceed 2 are counted modulo 2.

The following theorem holds
\begin{theorem}\label{th1}
Suppose that $t\in[0,T)$, where $T$ is the maximal value of $t$
such that \eqref{s2:1}-\eqref{s2:1a} has a unique solution. Then

\emph{(i)} The current means of $\varphi_i(\omega_i),\, i=1,2$,
are determined by the formulas
\begin{equation}\label{s2:3}
    \bar{\varphi}_i(t)=\left.\frac{dM_i(0,\lambda)}{d\lambda}\right|_{\lambda=q_i(t)}\frac{1}{M_i(0,q_i(t))}\,,
\end{equation}
and satisfy the equations
\begin{equation}\label{s2:4}
    \frac{d}{dt}\bar{\varphi}_i(t)=g_i(N_1,N_2,\bar{\varphi}_{i+1}(t))\sigma_i^2(t),
\end{equation}
where $\sigma_i^2(t)$ are the current variances of
$\varphi_i(t,\omega_i)$, $i=1,2$.

\emph{(ii)} The current population sizes $N_1(t)$ and $N_2(t)$
satisfy the system
\begin{equation}\label{s2:5}
    \frac{d}{dt}N_i(t)=N_i(t)[f_i(N_1,N_2,\bar{\varphi}_{i+1}(t))+\bar{\varphi}_i(t)g_i(N_1,N_2,\bar{\varphi}_{i+1}(t))],\quad
    i=1,2,
\end{equation}
where indexes that exceed 2 are counted modulo 2.
\end{theorem}

Theorem \ref{th1} gives a method of computation of the main
statistical characteristics of $\varphi_i(\omega_i)$; the analysis
of model \eqref{s2:1}-\eqref{s2:1a} is reduced to analysis of ODE
system \eqref{s2:2},\eqref{s2:3},\eqref{s2:5}, the only thing we
need to know is the mgfs of the initial distributions. It is worth
noting that the evolution of distributions can also be analyzed
\cite{Karev2005a}.

Concluding this sections we note that, with obvious notation
changes, models \eqref{s1:1}-\eqref{s1:3} and
\eqref{s1:6}-\eqref{s1:7} fall into the general framework of the
master model \eqref{s2:1}.

\section{Homogeneous models with nonlinear transmission
functions}\label{s4}
\subsection{Reduction to a system of ODEs}\label{s31}
We start with the equation \eqref{s1:1}, other equations in the
system can be quite arbitrary, e.g., the full system can contain
the class of exposed or several infective classes. Here we show
that the heterogeneous model that contains \eqref{s1:1} as a
modeling ingredient can be reduced to a homogeneous model with a
nonlinear transmission function whose explicit form is determined
by the initial distribution. This result is close to the analysis
in \cite{Veliov2005} where it was argued that the model with
heterogeneities can be encapsulated in a homogeneous model. Due to
the fact that the model considered in \cite{Veliov2005} is
substantially more general than the models we consider here, no
explicit formulas were found. In the case of model \eqref{s1:1}
nonlinear transmission function can be found in the exact form for
different initial distributions.

According to Theorem \ref{th1} we can rewrite equation
\eqref{s1:1} in the form
\begin{equation}\label{s3:1}
    \begin{split}
    \frac{d}{dt}S(t) &= -\bar{\beta}(t)S(t)I(t),\quad S(0)=S_0,\\
        &\ldots\\
        \frac{d}{dt}q(t)&=-I(t),\quad q(0)=0,\\
        \bar{\beta}(t)&=\left.\frac{dM(0,\lambda)}{d\lambda}\right|_{\lambda=q(t)}\frac{1}{M(0,q(t))},
\end{split}
\end{equation}
where dots denote equations that govern the dynamics of other
subpopulations, e.g., these can be usual equations for infected
and removed classes. $M(0,\lambda)$ is the given mgf of
$p_s(0,\omega)$.

\begin{proposition}\label{pr1}
Model \eqref{s3:1} is equivalent to the following model:
\begin{equation}\label{s3:2}
    \begin{split}
    \frac{d}{dt}S(t) & =-h(S(t))I(t),\\
        & \ldots
\end{split}
\end{equation}
where dots denote the same as in \eqref{s3:1},
\begin{equation}\label{s3:3}
    h(S)=S_0\left[\left.\frac{dM^{-1}(0,\xi)}{d\xi}\right|_{\xi=S/S_0}\right]^{-1},
\end{equation}
and $M^{-1}(0,\xi)$ is the inverse function to mgf $M(0,\lambda)$.
\end{proposition}
\begin{proof}
The first equation in \eqref{s3:1} can be rewritten in the form
$$
\frac{1}{S(t)}\frac{d}{dt}S(t)=\bar{\beta}(t)\,\frac{d}{dt}q(t)\,.
$$
From \eqref{s2:3} $\bar{\beta}(t)$ can be represented as $
\bar{\beta}(t)=\left.\frac{d\ln
M(0,\lambda)}{d\lambda}\right|_{\lambda=q(t)}, $ which, together
with the previous, gives
$$
\frac{d\ln S(t)}{dt}=\frac{d}{dt}\ln M(0,q(t)),
$$
or, using the initial conditions $S(0)=S_0,\,q(0)=0$,
\begin{equation}\label{s3:4}
    S(t)/S_0=M(0,q(t)),
\end{equation}
which is the first integral to system \eqref{s3:1}. Knowledge of a
first integral allows to reduce the order of the system by one.
Since $M(0,\lambda)$ is an absolutely monotone function in the
case of nonnegative $\beta(\omega)\geqslant 0$, then it follows
that
\begin{equation}\label{s3:5}
q(t)=M^{-1}\left(0,S(t)/S_0\right),
\end{equation}
where $M^{-1}(0,M(0,\lambda))=\lambda$ for any $\lambda$.

Putting \eqref{s3:5} into \eqref{s3:1} gives
$$
\frac{d}{dt}S(t)=\left.\frac{dM(0,\lambda)}{d\lambda}\right|_{\lambda=M^{-1}\left(0,S(t)/S_0\right)}S_0I(t),
$$
or, by the inverse function theorem, \eqref{s3:2} with
\eqref{s3:3}.
\end{proof}

The simple properties of the nonlinear incidence function $h(S)$
are $h(0)=0,\,h(S_0)=S_0\bar{\beta}(0),\,h'(S)>0$.

Let us consider several examples. Definitions for the probability
distributions we use can be found in Appendix. In all examples it
is assumed that $\beta(\omega)=\omega$, i.e., the transmission
coefficient takes the values from the domain of $\omega$ with the
probability corresponding to $p_s(t,\omega)$.

If the initial distribution is a gamma-distribution with
parameters $k$ and $\nu$ we obtain
\begin{equation}\label{gd1}
h(S)=\frac{kS}{\nu}\left[\frac{S}{S_0} \right]^{1/k}.
\end{equation}

If $k=1$ (i.e., the initial distribution is exponential with mean
$1/\nu$), then $h(S)=S^2/(\nu S_0)$. Expression \eqref{gd1} was
first obtained in \cite{Dwyer1997} and later used as a nonlinear
incidence function in~\cite{Dwyer2002}.

\begin{figure}[bth!]
\centering
\includegraphics[width=0.8\textwidth]{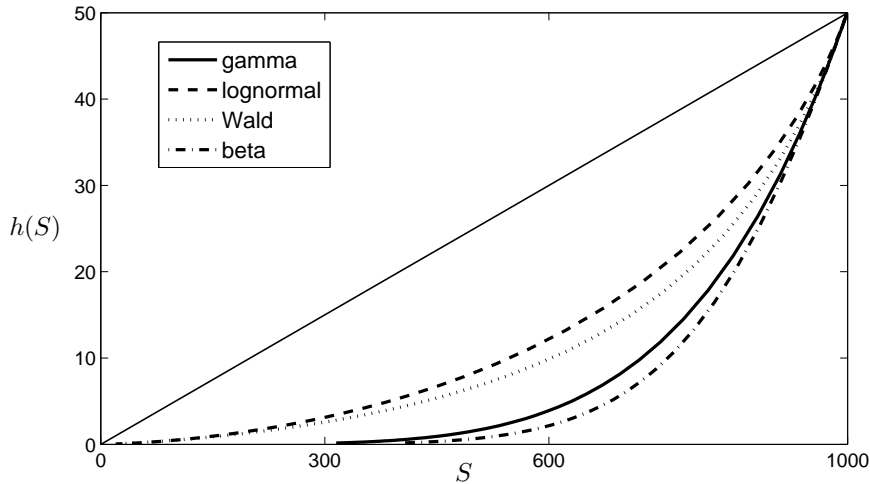}
\caption{The functions $h(S)$ given by \eqref{s3:3} for four
different probability distributions with the same means and
variances. The diagonal shows the same function for the
homogeneous model $h(S)=\beta S$}\label{f2}
\end{figure}

If the initial distribution is an inverse gaussian (Wald)
distribution with parameters $\mu$ and $\nu$, then
\begin{equation}\label{gd3}
    h (S) =\nu
    S\left[\frac{\nu}{\mu}-\ln\left(\frac{S}{S_0}\right)\right]^{-1}.
\end{equation}
    In a similar vein other initial distribution can be analyzed, but
not all of them have an explicit expression for mgf. Other
examples of possible initial distributions are uniform
distribution with parameters $a$ and $b$, lognormal distribution,
beta-distribution, Weibull distribution, Pareto distribution and
many others (see Fig. \ref{f2} for four different probability
distributions).

The theory we briefly presented in Section \ref{s2} is equally
valid both for continuous and discrete distribution. For instance,
if the initial distribution is the Poisson distribution with
parameter $\theta$, then we obtain
\begin{equation}\label{gd2}
  h(S)  =S\left[ \theta+\ln  \left( {\frac {S}{{\it S_0}}} \right)
  \right].
\end{equation}
Note that in the case of the Poisson distribution there is
non-zero probability that a randomly chosen individual has zero
transmission coefficient, i.e., the Poisson distribution implies
that some individuals are immune.

\subsection{Derivation of the power transmission function}

In standard epidemiological models the incidence rate (the number
of new cases in a time unit) was frequently used as a bilinear
function of infective and susceptible populations: $\propto SI$.
In addition to this it is usually argued that there are a variety
of reasons that the standard bilinear form may require
modification, including the assumption of heterogeneous mixing
\cite{Liu1987,roy2006}. We refer to a review paper on the subject
\cite{mccallum2001} for a general account of different models for
incidence rates, while noting that one of the most widely used
models has the form
\begin{equation}\label{in1}
\beta S^pI^q,\quad p,\,q>0,
\end{equation}
and is of direct interest to our study. The incidence rate in the
form \eqref{in1} was first used in \cite{severo1967}, with the
restriction that $p,\,q<1$, but generally it is only required that
$p,q>0$, see also \cite{Liu1986a,Hochberg1991,Knell1996}. It is
interesting to note in the context of our exposition that the
exponents $p,\,q$ in \eqref{in1} were dubbed as ``heterogeneity
parameters,'' but the models itself is considered phenomenological
and lacking mechanistical derivation \cite{mccallum2001}.

A special case of \eqref{in1} is $q=1$, which was considered in
\cite{Stroud2006,haas1999}; the values for parameter $p$ were
considered $p>1$. Comparison of the model \eqref{in1} with the ODE
\eqref{s3:2} when the initial distribution of the susceptible
subpopulation is a gamma-distribution (see \eqref{gd1}) let us
state the following corollary.

\begin{corollary}
The power relationship \eqref{in1} of the incidence rate in an SIR
model for the case $q=1,p>1$ can be obtained as a consequence of
the heterogenous model \eqref{s1:1}-\eqref{s1:3} when the initial
distribution of the susceptible subpopulation is a
gamma-distribution.
\end{corollary}

Let us assume that not only the susceptibles are heterogeneous for
some trait that influences the disease evolution, but also the
infectives are heterogeneous, and consider the simplest possible
SI model. Let $s(t,\omega_1)$ and $i(t,\omega_2)$ be the densities
of the susceptibles and infectives respectively, here we assume
that the traits of the two classes are independent, i.e.,
$\beta(\omega_1,\omega_2)=\beta_1(\omega_1)\beta_2(\omega_2)$. The
number of susceptibles with the trait value $\omega_1$ infected by
individuals with trait value $\omega_2$ is given by
$\beta_1(\omega_1)s(t,\omega_1)\beta_2(\omega_2)i(t,\omega_2)$,
and the total change in the infective class with trait value
$\omega_2$ is
$\beta_2(\omega_2)i(t,\omega_2)\int_{\Omega_1}\beta_1(\omega_1)s(t,\omega_1)\,d\omega_1$;
an analogous expression applies to the change in the susceptible
population. Combining the assumptions we obtain the following
model:
\begin{equation}\label{si1}
    \begin{split}
    \frac{\partial}{\partial t}s(t,\omega_1)  &=-\beta_1(\omega_1)s(t,\omega_1)\int_{\Omega_2}\beta_2(\omega_2) i(t,\omega_2)\, d\omega_2=-\beta_1(\omega_1)s(t,\omega_1)\bar{\beta}_2(t)I(t)\\
    \frac{\partial}{\partial t}i(t,\omega_2)  &=\beta_2(\omega_2)i(t,\omega_2)\int_{\Omega_1}\beta_1(\omega_1) s(t,\omega_1)\,
    d\omega_1=\beta_2(\omega_2)i(t,\omega_2)\bar{\beta}_1(t)S(t).
\end{split}
\end{equation}
Model \eqref{si1} is supplemented with initial conditions
$s(0,\omega_1)=S_0p_s(0,\omega_1),\,i(0,\omega_2)=I_0p_i(0,\omega_2)$.
In \eqref{si1} it is assumed that if an individual having trait
value $\omega_1$ was infected by an individual with trait value
$\omega_2$ he or she becomes an infective with trait value
$\omega_2$ (see Section \ref{s22}). The global dynamics of
\eqref{si1} is simple and is similar to the simplest homogeneous
SI model.

According to Theorem \ref{th1} the system \eqref{si1} can be
reduced to a four-dimensional system of ODEs. Reasoning exactly as
in the proof of Proposition \ref{pr1} we obtain

\begin{proposition}\label{pr2}
The model \eqref{si1} is equivalent to the model
\begin{equation*}
    \begin{split}
    \frac{d}{dt}S(t) &= -h_1(S)h_2(I),\\
    \frac{d}{dt}I(t) &= h_1(S)h_2(I),
\end{split}
\end{equation*}
where $h_i(x)$ is given by \eqref{s3:3}.
\end{proposition}
Combining Proposition \ref{pr2} with \eqref{gd1} we get

\begin{corollary}
The power relationship \eqref{in1} of the incidence rate in an SI
model for the case $q>1,p>1$ can be obtained as a consequence of
the heterogenous model \eqref{si1} when the initial distributions
of the populations of susceptibles and infectives are
gamma-distributions.
\end{corollary}

Consequently, it turns out that the power relationship
\eqref{in1}, at least for the case $p,\,q\geqslant 1$, can be
explained on the mechanistic basis by the inherent heterogeneities
of the population. Its exact form is the consequence of the
initial gamma-distributions, but we note that any of the
transmission functions given in the previous subsection can be
well approximated by \eqref{gd1} (see also Fig. \ref{f2}).

\section{The influence of population heterogeneity on the disease
course}\label{s5} Here we mainly restrict our attention to the
model \eqref{s1:1}-\eqref{s1:3} and study its global behavior.
First we state almost obvious proposition:

\begin{proposition}\label{pr3}
Let $S_1(t)=\int_{\Omega}s_1(t,\omega)\,d\omega$ be the solution
of \eqref{s1:1}-\eqref{s1:3} with the initial condition
$s_1(0,\omega)=S_0p_1(0,\omega)$, and
$S_2(t)=\int_{\Omega}s_2(t,\omega)\,d\omega$ be the solution of
\eqref{s1:1}-\eqref{s1:3} with the initial condition
$s_2(0,\omega)=S_0p_2(0,\omega)$, such that
$\bar{\beta}_1(0)=\bar{\beta}_1(0)$ and
$\sigma^2_1(0)>\sigma^2_1(0)$, where
$\bar{\beta}_i(0)=\int_{\Omega}\beta(\omega)p_i(0,\omega)\,d\omega$
and
$\sigma^2_i(0)=\int_{\Omega}(\beta(\omega)-\bar{\beta}_i(0))^2p_i(0,\omega)\,d\omega$,
$i=1,2$. Then there exists $\varepsilon>0$  such that
$S_1(t)>S_2(t)$ for all $t\in(0,\,\varepsilon)$.
\end{proposition}
The gist of this proposition is very simple: the more
heterogeneous the susceptible hosts the less severe the disease
progression under the model \eqref{s1:1}-\eqref{s1:3}.
\begin{proof}
Differentiating the first equation in \eqref{s3:1} and using
\eqref{s2:4} we get
$$
S''(t)=I^2S(\sigma^2(t)+\bar{\beta}(t))-\bar{\beta}(t)I'S,
$$
or, at the initial time moment, $S_1''(0)>S_2''(0)$. Since $S(t)$
is continuous the proposition follows.
\end{proof}
We remark that this proposition also holds for more general model
\eqref{s3:1}. Moreover, we can replace equation \eqref{s1:1} with
equation
$$
\frac{\partial}{\partial
t}s(t,\omega)=-\beta(\omega)s(t,\omega)I(t)+s(t,\omega)[\ldots]
$$
where dots denote terms describing demography, migration or the
lost of immunity by removed individuals, the only condition is
that these terms cannot depend on $\beta(\omega)$. Even in this
case Proposition \ref{pr3} still holds. For the model
\eqref{s1:5}-\eqref{s1:6} the opposite proposition is true:
the more heterogeneous the infective class in infectivity, the
more severe the disease progression, which follows from the fact
that $\bar{\beta}'(t)>0$ in this case, and, consequently,
$S_1''(0)<S_2''(0)$.

It is interesting to note that in the case of proportional mixing
(frequency-dependent transmission) knowledge of only the initial
variances of the parameter distributions does not allow inference
on the short ran behavior \cite{Veliov2005}.

One of the main characteristic of SIR models is the final size of
the disease, which is often expressed in the number (or
proportion) of susceptibles that never get infected. For the
Kermack-McKendrick model
$$dS/dt=-\beta SI,\quad dI/dt=\beta SI-\gamma I,\quad dR/dt=\gamma I,$$ it is well known
that the desired number, which we denote as $S(\infty)$, is given
by the root of the equation
\begin{equation}\label{fs1}
S(\infty)=S_0\exp\left\{-\frac{\beta}{\gamma}(N-S(\infty))\right\}
\end{equation}
on the interval $(0,S_0)$. Here $N$ is the constant size of the
population. It is easy to show that this root always exists.

Recall that $M(0,\lambda)$ is the mgf of the initial parameter
distribution. For the model \eqref{s1:1}-\eqref{s1:3} the
following theorem holds.

\begin{theorem}
The size of the susceptible subpopulation that escapes infection
within the framework of the model \eqref{s1:1}-\eqref{s1:3} is
given by the solution of the equation
\begin{equation}\label{f_s}
    S(\infty)=S_0M\left(0,(S(\infty)-N)\gamma^{-1}\right)
\end{equation}
satisfying condition $0<S(\infty)<S_0$.
\end{theorem}
\begin{proof}
First we note that exactly as it was done in Proposition
\ref{pr1}, we can reduce the system \eqref{s1:1}-\eqref{s1:3} to
the system
\begin{equation}\label{n2}
\begin{split}
    dS(t)/dt &= -h(S(t))I(t),    \\
    dI(t)/dt &= h(S(t))I(t)-\gamma I(t),\\
    dR(t)/dt&=\gamma I(t).
\end{split}
\end{equation}
Using \eqref{s3:3} and dividing the first equation in \eqref{n2}
by the third one we obtain
$$
\left.\frac{dM^{-1}(0,\xi)}{d\xi}\right|_{\xi=S/S_0}\frac{dS}{S_0}=-\gamma^{-1}dR.
$$
Integrating from $0$ to $\infty$ gives
$$
\int_{1}^{S(\infty)/S_0}dM_0^{-1}(\xi)=-\frac{R(\infty)}{\gamma}\,.
$$
Using the identities $R(\infty)=N-S(\infty)$ (since $I(\infty)=0$)
and $M^{-1}(0,1)=0$ we obtain
$$
M_0^{-1}(S(\infty)/S_0)=-\frac{N-S(\infty)}{\gamma}\,,
$$
from which \eqref{f_s} follows. Due to the fact that
$M(0,\lambda)$ is an increasing function, the solution of
\eqref{f_s} satisfying $0<S(\infty)<S_0$ is unique.
\end{proof}

\textit{Remark.} If we consider nondistributed parameter
(formally, we can let $\beta(\omega)=\beta=\mbox{const}$, or,
equivalently,
$s(0,\omega)=S_0\delta(\omega-\bar{\omega}),\,\beta(\bar{\omega})=\mbox{const}$,
where $\delta(\omega)$ is the delta-function), we obtain
\eqref{fs1} from \eqref{f_s}.

Arguing in the same spirit as it is done in \cite{Diekmann2000}
(e.g., p. 183), the problem of the epidemic invasion can be
considered. Assuming that initially all the population is
susceptible (formally, for our model,
$S(-\infty)=N,\,q(-\infty)=0$), from \eqref{f_s} the equation for
the fraction of susceptible population that does not get infected
follows:
\begin{equation}\label{epin}
    z=M(0,-N(1-z)/\gamma).
\end{equation}
Here $z=S(\infty)/N$. Equation \eqref{epin} always has the root
$z=1$. If the basic reproductive number \cite{Diekmann1990},
defined here as
\begin{equation}\label{r0}
    R_0=\frac{\bar{\beta}(0)N}{\gamma},
\end{equation}
satisfies the condition $R_0> 1$, then there is another root of
\eqref{epin} in the interval $0<z<1$. This root gives the sought
fraction. The proof of the existence of this root under the
threshold condition $R_0>1$ is straightforward and can be
conducted similar to the homogeneous case
(e.g.,~\cite{Diekmann2000}). This result can be illustrated by the
case when the initial distribution is exponential with parameter
$\nu$: the equation for $z$ is quadratic: $R_0z^2-(1+R_0)z+1=0$,
where $R_0=N/(\gamma\nu)$. This equation has the roots $1$ and
$1/R_0$. If $R_0>1$ then the fraction of susceptible population
that escapes the disease is $1/R_0$.

Comparing the results obtained for the heterogeneous SIR model
\eqref{s1:1}-\eqref{s1:3} with the well known results for the
simple homogeneous SIR model, we can conclude that the questions
of the disease invasion can be studied in the framework of the
homogeneous model because population heterogeneity does not impact
the basic reproductive number \eqref{r0} (this holds, obviously, if we identify the mean
value of $\beta(\omega)$ over the population of susceptibles at
the initial time moment with the usual constant $\beta$ in the
homogeneous model). From the other hand, the heterogeneity of the
population has direct impact on the final size of the disease
since equation \eqref{epin} depends on the initial distribution in
contrast to the homogeneous analogue $z=\exp\{-R_0(1-z)\}$
(surprisingly, the last formula is  valid under variety of
different conditions \cite{ma2006}).
\begin{figure}[bth!]
\centering
\includegraphics[width=0.8\textwidth]{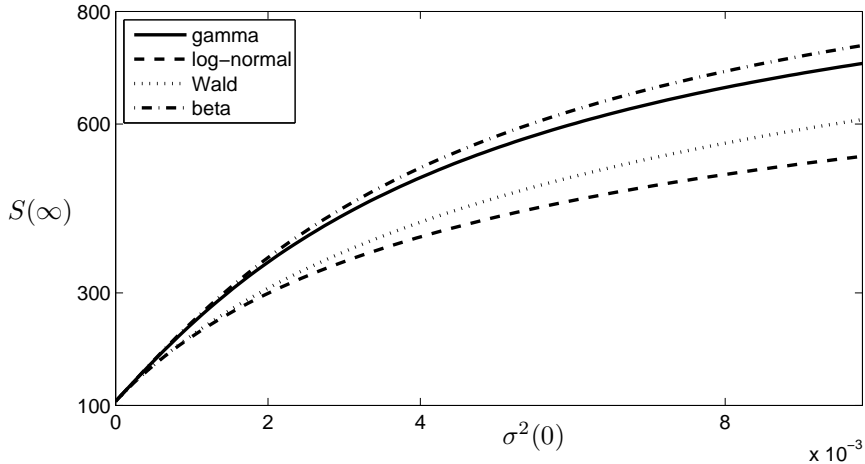}
\caption{The size of the susceptible population that never gets
infected, $S(\infty)$ versus the initial variance of the parameter
distribution $\sigma^2(0)$ for four different initial
distributions, $\bar{\beta}(0)$ is the same in all cases. The
parameters are
$S(0)=999,\,I(0)=1,\,\gamma=20,\,\bar{\beta}(0)=0.05$} \label{f1}
\end{figure}

In Fig. \ref{f1} the final size of the susceptible population
versus the initial variance of the parameter distribution is shown
for four different initial distributions that have the same means
at $t=0$. From Fig. \ref{f1} it can be seen that the more
heterogeneous the population of susceptibles, the less severe the
disease not only in the short run (Proposition \ref{pr3}) but also
globally (see \cite{May1988} for the same result for
frequency-dependent transmission). At the same time, it is worth
emphasizing that the conditions
$\bar{\beta}_1(0)=\bar{\beta}_2(0),\,\sigma^2_1(0)>\sigma^2_2(0)$
for two different  initial distributions do not imply that
$z_1>z_2$, where $z_i$ are the solutions of \eqref{epin}. A
counterexample can be easily found (e.g., taking
gamma-distribution with parameters $k=2,\,\nu=4$ and uniform
distribution on the interval $[0,1]$, $N/\gamma=20$, we find that
$z_1=0.093<z_2=0.112$ whereas $\sigma_1(0)=1/8>\sigma_2(0)=1/12$).

If we compare two distributions of the same family then sometimes
it is possible to prove rigorously that, on the assumption of
equal initial means and different second moments, more
heterogeneous population (i.e., the one that has larger initial
variance) experiences less severe disease. This is true, e.g., for
two gamma-distributions:

\begin{proposition}
Assume that we have model \eqref{s1:1}-\eqref{s1:3} with two
initial gamma-distributions with parameters $k_1,\nu_1$ and
$k_2,\nu_2$ such that $\bar{\beta}_1(0)=\bar{\beta}_2(0)$ and
$\sigma_1^2(0)>\sigma^2_2(0)$. Assume that $R_0>1$. Then solutions
of \eqref{epin} that belong to $(0,1)$ are always satisfy
$z_1>z_2$.
\end{proposition}
\begin{proof}
We need to prove that
$$
\frac{(k_2+R_0(1-z))^{k_2}}{(k_1+R_0(1-z))^{k_1}}\frac{k_1^{k_1}}{k_2^{k_2}}>1
$$
for any $z\in(0,1)$. This follows from the fact that
$f(r)=(k_2+r)^{k_2}/(k_1+r)^{k_1}$ is monotonically increasing
function for $r>0$, i.e., $f'(r)>0$.
\end{proof}

\textit{Remark.} The last proposition can be extended to other
initial distributions, e.g., it holds for Wald distribution, for a
uniform distribution and some others. However, for any
distribution that is not determined by its first two moments
(i.e., that depends on more than two parameters) an analogous
proposition is no longer true.

\section{Discussion and Conclusions}
Here we have presented several results concerning the course of a
disease in a closed heterogeneous population, where heterogeneity
is mediated by invariable traits whose distributions are
determined by population structure at every time moment. One of
the purposes of the present text is to introduce in the area of
epidemiological modeling the general technics of the theory of
heterogeneous populations \cite{Karev2000a,Karev2005a} which
allows reduction of the initial infinite-dimensional dynamical
system to an ODE system of a low dimension.

A usual strategy in the literature when analyzing systems similar
to \eqref{s1:1}-\eqref{s1:3} is to consider an
infinite-dimensional system of ODEs where the variables can be
moments or cumulants of the corresponding distributions (e.g.,
\cite{Dwyer2000,Nikolaou2006,Veliov2005}) and then analyze this
system, or consider one of many possible moment-closure strategies
to extract valuable information. Having the general theory
outlined in Section \ref{s2} and the known results from the
literature, we can critically discuss the latter and be prepared
to possible pitfalls.

For example, in \cite{Dwyer2000} the equation for the final
epidemic size was obtained by using two different strategies:
first, an initial gamma-distribution was assumed and analytical
treatment was applied, and second, using the procedure suggested
in \cite{Dushoff1999}, an approximation method was used in which
the infinite-dimensional system of ODEs was replaced with two
equations under the assumption that the coefficient of variation
is constant. It is not surprising that Dwyer et al. obtained
identical results because the gamma-distribution, according to the
theory of heterogeneous populations, is the only continuous
distribution which does not change its shape during the system
evolution, and keeps the coefficient of variation constant (see
Appendix). Therefore, the conclusion that ``...the assumption of
gamma-distributed susceptibility is not strictly necessary to
derive equation [for the final epidemic size]'' is not valid in
many situations. Another initial distribution, how it is shown by
\eqref{f_s}, can yield another equation for the final epidemic
size and, consequently, can produce significant discrepancy with
the moment-closure approximation suggested by Dushoff
\cite{Dushoff1999}.

The well known fact from the theory of heterogeneous populations
that to model the system dynamics for a substantial time period we
need to know the \textit{exact} initial distribution implies that
any results obtained for an epidemic in a heterogeneous population
on the ground of knowledge of only several first moments of the
initial distribution have to taken with extreme care. One, two, or
more first moments of the initial distribution can be insufficient
or even misleading. We also note that the short run behavior can
be predicted when we have information only on several first
moments (Section \ref{s5}).

Another initial distribution which was used in the literature is
the normal distribution \cite{Nikolaou2006}. For the normal
distribution we have that $\kappa_i=0,\,i\geqslant3,$ where
$\kappa_i$ is the $i$-th cumulant. Combining the last property
with the fact that the initial normal distribution remains normal
within the framework of heterogeneous models we obtain that
$\kappa_i(t)=0,\,i\geqslant3,$ holds for any $t$ (see Appendix).
This was used in \cite{Nikolaou2006} to obtain an explicit
solution to the equation
$$
\frac{\partial}{\partial t}n(t,r)=K_gn(t,r)-rn(t,r).
$$
Here $n(t,r)$ is the number of cells at time moment $t$, which are
killed by antimicrobial agents with the kill rate $r$, and $K_g$
is a constant (notations are changed from the original). First we
note that, using Theorem \ref{th1}, we obtain explicit solution of
this equation for an arbitrary initial distribution of the kill
rate:
$$
N(t)=N_0M(0,-t)\exp\{K_gt\},
$$
where $N(t)=\int_{R}n(t,r)\,dr$. Second, the results from Sections
\ref{s4} and \ref{s5} cannot be applied to the normal distribution
because this distribution is defined from $-\infty$ to $\infty$
and thus the corresponding mgf does not have an inverse. Which is
more important, however, the total system dynamics can be
influenced by these negative kill rates even if they occur with
vanishingly small probability (we note that this issue is
discussed in \cite{Nikolaou2006}). An example of such influence
can be found in \cite{Karev2005} where infinitely large growth
rates occurring with small probabilities drive the population to
explosion. Therefore any approximations based on an initial normal
distribution in a situation where parameter can take only
nonnegative values should be taken with care.

Summarizing the main results we can assert that the theory of
heterogeneous populations can be successfully applied to many
different mathematical models in epidemiology. Examples are given
in Section \ref{s2}. In many simple cases the original model can
be reduced to a model described by ODEs, which simplifies the
analysis. The law of mass action for a distributed susceptibility
model implies a nonlinear incidence function in a homogeneous
model. Moreover, one of the well known transmission functions,
power relationship, follows in exact form from the initial
gamma-distribution, at least in the case when exponents exceed one
(see Section \ref{s4}). Therefore, a mechanistic derivation has been given to the transmission power function, which was shown previously approximate real data with high accuracy. The short term behavior of the models
considered can be approximately described knowing only two first
moments of the initial distribution, whereas the long-term
behavior depends on the exact initial distribution and can vary
significantly (Section \ref{s5}) even for the distributions whose
several first moments are identical (Fig. \ref{f1}).

It is a tempting challenge to include various demography processes
to the analyzed models. The main obstacle is the need to specify
the function $\phi(\omega,\omega')$ similar to the one used in
$\eqref{s1:5}$. The delta-function yields the models that can be
analyzed using the general approach from Section \ref{s2} but it
is usually difficult to interpret the underlying assumptions.
These problems are the subject of ongoing research.

\begin{appendix}
\section{Appendix}
In Appendix we collect the definitions of the distributions used
throughout the main text. The definitions are taken from
\cite{Johnson1994} and \cite{Johnson1995}. In addition to that we
list some facts concerning evolution of these distributions if
they are used as the initial distributions for the models studies
in the text. Everywhere below it is assumed that
${\beta}(\omega)=\omega$. Inasmuch as we are interested in
characteristics of distributions depending on time, the following
formula is very useful (see \cite{Karev2005a}):
\begin{equation}\label{ap1}
    M(t,\lambda)=\frac{M(0,\lambda+q(t))}{M(0,q(t))},
\end{equation}
where $q(t)$ in the solution of the corresponding auxiliary
differential equation (see Theorem \ref{th1}), $M(t,\lambda)$ is
the mgf of the parameter distribution at time $t$. Equation
\eqref{ap1} shows that the mgf at any time instant can be
expressed using the initial mgf.

\textit{Gamma-distribution}

The pdf of gamma-distribution with parameters $k$ and $\nu$ is
given by
\begin{equation}\label{gamma}
    p(0,\omega)=\frac{\nu^k}{\Gamma(k)}\omega^{k-1}e^{-\nu\omega},\quad
    \omega\geqslant 0,\,k>0,\,\nu>0.
\end{equation}
The mgf of gamma-distribution is
$M(0,\lambda)=(1-\lambda/\nu)^{-k}$.

It follows from \eqref{ap1} that for $t>0$ the distribution does
not change its form, i.e., it is gamma-distribution with
parameters $k$ and $s-q(t)$. The mean and variance of the
distribution are given by
$$
\bar{\beta}(t)=\frac{k}{s-q(t)}\,,\quad
\sigma^2(t)=\frac{k}{(s-q(t))^2}\,.
$$
Note that at any time moment the coefficient of variation is
constant: $cv=\sigma(t)/\bar{\beta}(t)=1/\sqrt{k}$. Actually,
gamma-distribution is the only continuous distribution whose
coefficient of variation remains constant with time within the
framework of heterogeneous models.

 \textit{Wald distribution}

The pdf of inverse gaussian (Wald) distribution with parameters
$\mu$ and $\nu$ is
\begin{equation}\label{d3}
    p(0,\omega)=\left[\frac{\nu}{2\pi
    \omega^3}\right]^{1/2}\exp\left\{-\frac{\nu}{2\mu^2\omega}(\omega-\mu)^2\right\},\quad
    \omega>0,\quad \mu,\,\nu>0.
\end{equation}
The mgf of Wald distribution is given by
$$
M(0,\lambda)=\exp\left\{\frac{\nu}{\mu}\left(1-\left[1-\frac{2\mu^2\lambda}{\nu}\right]^{1/2}\right)\right\}.
$$
Again the distribution remains Wald distribution with parameters
$$
\mu\left[1-\frac{2\mu^2q(t)}{v}\right]^{-\frac{1}{2}},\,\nu,
$$
and temporal characteristics of the distribution are
$$
\bar{\beta}(t)=\mu\left[1-\frac{2\mu^2q(t)}{v}\right]^{-\frac{1}{2}},\quad
\sigma^2(t)=\frac{\bar{\beta}^3(t)}{\nu}\,,\quad
cv^2=\frac{\bar{\beta}(t)}{\nu}\,.
$$

\textit{Beta-distribution}

Sometimes it is useful to study evolution of distribution with
compact support. A good candidate in this case is the family of
beta-distributions with pdf
\begin{equation}\label{d4}
    p(0,\omega)=\frac{1}{B(r_1,\,r_2)}\frac{(\omega-a)^{r_1-1}(b-\omega)^{r_2-1}}{(b-a)^{r_1+r_2-1}}\,,\quad
    a\leqslant\omega\leqslant b,\,r_1>0,\,r_2>0,
\end{equation}
where $B(r_1,r_2)$ is the beta-function.

The initial mean and variance are
$$
\bar{\beta}(0)=\frac{r_1}{r_1+r_2},\quad
\sigma^2(0)=\frac{r_1r_2}{(r_1+r_2)^2(r_1+r_2+1)}\,.
$$
Unfortunately in the case of beta-distribution it is impossible to
write down the mgf, and, correspondingly, the temporal
characteristics of the distribution. In a special case $r_1=r_2=1$
we have a uniform distribution on $[a,\,b]$. Equation \eqref{ap1}
shows that in this case for $t>0$ the distribution is no longer
uniform but turns into truncated exponential distribution.

In the text we used beta-distribution on $[0,1]$.

\textit{Log-normal distribution}

The pdf of log-normal distribution defined on the nonnegative
half-axis $\omega\geqslant 0$ with parameters $\mu$ and $\nu$ is
\begin{equation}\label{d5}
    p(0,\omega)=\left[\omega\sqrt{2\pi}\nu\right]^{-1}\exp\left\{-\frac{1}{2}\frac{(\ln\omega-\mu)^2}{\nu^2}\right\},\quad
    \omega\geqslant0,\,\mu>0,\,\nu>0.
\end{equation}
The initial characteristics are
$$
\beta(0)=\exp\{\mu+\nu^2/2\},\quad
\sigma^2(0)=\exp\{2\mu+\nu^2\}(\exp\{\nu^2\}-1).
$$
As in the case of beta-distribution we cannot present explicit
formulas for the mgf and other characteristics. We note that this
distribution can be used only if $q(t)<0$ for $t>0$, otherwise the
integral in the mgf diverges. This is the case, e.g., for the
model \eqref{s1:1}-\eqref{s1:3}, but not the case for
\eqref{s1:6}-\eqref{s1:7}, for which the log-normal distribution
cannot be used.

\textit{Normal distribution}

The pdf is
\begin{equation}\label{d6}
    p(0,\omega)=\left[\sqrt{2\pi}\sigma\right]^{-1}\exp\left\{-\frac{(\omega-\mu)^2}{2\sigma^2}\right\},\quad
    \sigma>0.
\end{equation}
The mgf is
$M(0,\lambda)=\exp\left\{\frac{1}{2}\lambda(2\mu+\lambda\sigma^2)\right\}$.
The temporal characteristics are
$$
\bar{\beta}(t)=\mu+2q(t)\sigma^2,\, \sigma^2(t)=\sigma^2.
$$

\textit{Poisson distribution}

In the same spirit discrete distributions can be managed.
Consider, e.g., the Poisson distribution with parameter $\theta$,
i.e,
\begin{equation}\label{d2}
    p(0,\omega)=\frac{\theta^\omega}{\omega!}e^{\theta},\quad
    \theta>0,\,\omega=0,\,1,\ldots
\end{equation}
then $M(0,\lambda)=\exp\left\{\theta(e^\lambda-1)\right\}$, and
$$
\bar{\beta}(t)=\sigma(t)=\theta\exp\{q(t)\},\, cv=1.
$$

Other possible initial distribution can be considered in a similar
vein.

\end{appendix}

\smallskip
\noindent \textbf{Acknowledgments.} The author thanks Dr. A.
Bratus' and Dr. G. Karev for insightful discussions and helpful
suggestions.

\end{document}